\documentclass[]{aesconf}

\graphicspath{{./}{figures/}}

\usepackage[utf8]{inputenc}
\usepackage[T1]{fontenc}
\usepackage{amsmath}
\usepackage{subcaption}
\usepackage{gensymb}
\usepackage{xcolor}
\usepackage{siunitx}

\usepackage{microtype}

\usepackage[numbers,square]{natbib}

\usepackage{booktabs}
\usepackage{color}
\usepackage{url}
\usepackage{graphicx}
\usepackage{siunitx}
\usepackage{overpic}
\usepackage{soul}

\newcommand{\DS}{\text{DS}}

\newcommand\thefont{\expandafter\string\the\font}
\title{Improved Panning on Non-Equidistant Loudspeakers with Direct Sound Level Compensation}

\author[ \hspace{-0.75ex}]{Jan-Hendrik Hanschke}
\author[ \hspace{-0.75ex}]{Daniel Arteaga}
\author[ \hspace{-0.75ex}]{Giulio Cengarle}
\author[ \hspace{-0.75ex}]{Joshua Lando}
\author[ \hspace{-0.75ex}]{Mark R.P. Thomas}
\author[1]{Alan Seefeldt}

\affil[1]{Dolby Laboratories}

\correspondence{Jan-Hendrik Hanschke}{janhendrikhanschke@ieee.org}

\lastnames{Hanschke et al.}

\shorttitle{Improved Panning on Non-Equidistant Loudspeakers with Direct Sound Level Compensation}

\begin{document}

\twocolumn[
\maketitle %

\begin{onecolabstract}

Loudspeaker rendering techniques that create phantom sound sources often assume an equidistant loudspeaker layout. 
Typical home setups might not fulfill this condition as loudspeakers deviate from canonical positions, thus requiring a corresponding calibration. 
The standard approach is to compensate for delays and to match the loudness of each loudspeaker at the listener's location.
It was found that a shift of the phantom image occurs when this calibration procedure is applied and one of a pair of loudspeakers is significantly closer to the listener than the other.
In this paper, a novel approach to panning on non-equidistant loudspeaker layouts is presented whereby the panning position is governed by the direct sound and the perceived loudness is governed by the full impulse response. 
Subjective listening tests are presented that validate the approach and quantify the perceived effect of the compensation. In a setup where the standard calibration leads to an average error of 10\degree, the proposed direct sound compensation largely returns the phantom source to its intended position.

\end{onecolabstract}
]

\section{Introduction}\label{sec:Introduction}

In stereo or multichannel loudspeaker setups, a virtual or phantom source is a sound that appears to emanate from a position other than the physical loudspeaker locations \cite{rumsey_12_spatial}. 
The most common rendering techniques for creating such phantom sources are based on stereo amplitude panning and their multichannel extensions (e.g., vector-base amplitude panning \cite{pulkki_97_vbap}, dual/triple balance amplitude panning \cite{thomas_17_panning}, distance-based amplitude panning \cite{lossius_09_dbap}). These panning methods distribute the source signal among several loudspeakers, assigning a gain to each loudspeaker so that the resulting sound mixture creates the illusion of a phantom sound source coming from the intended direction. Amplitude panning techniques are commonly used in professional content creation tools for cinema, music and multimedia. 

With traditional channel-based formats, panning %
to channels takes place at the content creation side, addressing a small discrete set of canonical playback configurations (e.g., stereo, 5.1, etc.). These channel-based renderings are then played back on consumer systems where the loudspeaker positions may deviate from the canonical locations, causing a mismatch in angle and perceived level.  These inaccuracies result in a shift of the perceived position of a phantom source with respect to the intended position. Object-based audio \cite{tsingos2017object}, which utilizes a renderer in the playback device and knowledge of the loudspeaker layout, opens the door to modifying relative gains of individual sources based on the knowledge of actual loudspeaker location and acoustic characteristics of the playback system.  So far, most rendering techniques, including those that allow for flexible positioning of loudspeakers and process object-based audio, depend only on the angular position of the loudspeakers relative to the listener. The distance between each loudspeaker and the listening position is assumed to be equal, even if in common home setups that might not hold true.

In case of unequal distances, the state of the art approach is to time align and loudness match the different loudspeakers \cite{pulkki_97_vbap}, with the loudness estimated from the full room response of each loudspeaker, which we will refer to as full response compensation (FRC).
In the authors' experience, this calibration approach fails when rendering content to layouts with non-equidistant loudspeakers, causing the phantom source to be systematically pulled towards the closest loudspeaker(s). Upon a more thorough reflection with regards to the position of a phantom source, the procedure of loudness matching seems to be at least partially at odds with the well established psychoacoustic principle of the Haas or precedence effect \cite{gardner_historical_2005, haas1951einflubeta}. When a sound is followed by a delayed version of itself with a time delay of approximately \SI{1}{ms} or more (but less than the echo threshold), a single auditory event is perceived from the direction of the first arriving wavefront. As a consequence, the perceived direction of a single physical sound source in a room is dominated by sound on the direct path from the source to the listener, not by later arriving room reflections \cite{blauert_acoustic_2005}. %
For time delays smaller than approximately \SI{1}{ms}, the related summing localization principle \cite{blauert_96_spatial} states that multiple wavefronts of sound fuse into a phantom source whose perceived direction is a combination of those for each wavefront. When considering panning across multiple time-aligned loudspeakers, this gives strong indication that the quantity determining the virtual source location is the direct sound from each loudspeaker, possibly including early reflections arriving before \SI{1}{ms}, and not the total sound loudness contained in the entire reverberation tail. 
Although some literature in the context of equalization hints at the possibility that the direct sound plays a dominant role in the localization and timbre perception of sound~\cite{bank_13_quasi-anechoic, cecchi_14_eq}, and another study uses the anechoic decay as a simplified room model in a panning function~\cite{Matthews2015}, %
we are not aware of systematic studies of this phenomenon in the context of loudspeaker rendering, nor of any practical implementations. %

We  study and propose a modified panning approach for non-equidistant loudspeakers based on the combined contributions of the direct sound from multiple loudspeakers, and show empirically that it leads to improved phantom source localization accuracy. In order to achieve loudness consistency across multiple phantom source locations, the total response of the loudspeakers is simultaneously considered.  %

The paper is organized as follows: In Sec.~\ref{sec:fundamentals} we introduce relevant quantities based on the sound decay model and explain the full response compensation method for level and delay compensation. Sec.~\ref{sec:DSC} then covers the proposed approach to restore the intended phantom source position based on the direct sound contribution while maintaining loudness consistency. In Sec.~\ref{sec:experiment} we describe a subjective listening test to validate the proposed approach and in Sec.~\ref{sec:results} we present its results. These results and the main outcomes of the paper are discussed in Sec.~\ref{sec:summary}.

\section{Fundamentals}\label{sec:fundamentals}
\subsection{Distance-based level decay of loudspeakers in reverberant rooms}\label{sec:distancedecaymodel}

As a loudspeaker plays a signal in a room, the direct sound---the sound traveling on the shortest path from the loudspeaker to the listener---is quickly followed by multiple, spatially diverse, indirect reflections with increasing temporal density, often referred to as the diffuse sound field. %

The direct sound intensity decays as the squared distance from the loudspeakers. The corresponding direct sound level for each loudspeaker, $L_i^\DS$ in decibel scale is
\begin{equation}\label{eq:Lds}
    L^\DS_i =10\, \log_{10} \left( \frac{P_i Q_i}{4\pi d_i^2} \right),
\end{equation}
where $P_i$ is the acoustic power of a source $i$, $Q_i$ its directivity factor in the direction of the listener, and $d_i$ is the distance to the source.

The diffuse sound intensity is almost constant, depends upon the room characteristics and varies little with source and receiver position, orientation, or distance. The distance at which direct and diffuse intensities are equal is commonly referred to as the critical distance $D_c$.
The overall loudness at the listening position for a given loudspeaker can be inferred by assuming the total sound as the sum of direct sound and diffuse sound field. %
On axis, the loudness can be estimated from the total sound intensity in decibel scale as
\begin{equation}
    L_i =10\, \log_{10} \left( \frac{P_i Q_i}{4\pi} \left[  \frac{1}{d_i^2} + \frac{1}{D_c^2}\right]  \right).
\end{equation}

In practice $L_i$ can also be obtained from the measurement with a sound level meter, e.g.~when capturing pink noise. %
An equivalent calibration can be achieved through the acquisition and analysis of impulse responses (IRs). The loudness can be estimated from the RMS value of the IR $h_i(t)$, with optional weighting filters applied $w_A (t)$, for example A-weighting:
\begin{equation}\label{eq:levelfromir}
    L_i = 10\, \log_{10} \left(\frac{1}{T} \int_0^T \left| (h_i \ast w_A) (t)\right|^2 dt\right).
\end{equation}

\begin{figure}[t]
\begin{center}
\includegraphics[width=0.95\columnwidth]{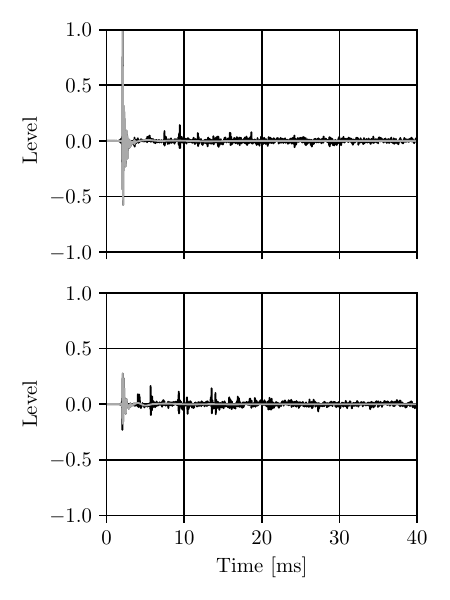}
\caption{Measured IR (black line) for a loudspeaker at \SI{1.5}{\meter} (top) vs.~\SI{3}{\meter} (bottom) distance, leading to a theoretical direct sound decay of \SI{6}{dB}. Analysis of IRs with Frequency Dependent Truncation (grey line) shows a \SI{6.3}{dB} level difference (pink- and A-weighted) in direct sound vs.~\SI{3.0}{dB} in overall sound (black line) between near and far loudspeaker.}
\label{fg:IRs}
\end{center}
\end{figure}

The direct sound level $L_i^\DS$ for each loudspeaker cannot be measured with a sound level meter. One option to obtain it is to use the room-independent model per ~\eqref{eq:Lds}. It can also be estimated by multiplying measured impulse responses with a time window that vanishes beyond a certain truncation time $\tau$ after the arrival of the first peak. Using a small, fixed truncation time has the drawback that frequencies approximately lower than the inverse truncation time cannot be adequately represented. A frequency-dependent truncation (FDT) kernel $k(n)$ \cite{karjalainen_2001_frequency} may be used to estimate the direct sound portion of the impulse response:
\begin{equation}
    h^\DS (t) = \int_0^\tau k(t,t') h(t') dt'.
\end{equation}
The frequency-dependent truncation filter truncates all frequency components of the impulse response to a time $\tau$ or smaller. Most commonly, it truncates the lowest frequency under consideration to a time $\tau$ and higher frequencies to a time smaller than $\tau$. This approach has the advantage of providing a better representation of the lower frequencies without compromising the truncation of the impulse response at higher frequencies.
Fig.~\ref{fg:IRs} shows examples of FDT applied to impulse responses of non-equidistant loudspeakers in a reverberant room.
The corresponding direct sound level $ L_i^\DS$ can be estimated by substituting $h_i$ for $h^\DS_i$ in \eqref{eq:levelfromir}.

\subsection{Full response compensation (FRC) of non-equidistant loudspeakers}\label{sec:SOTACalib}

The state of the art calibration approach involves loudness matching and loudspeaker time alignment.

Loudness matching ensures that each loudspeaker produces the same loudness at the listening position when fed with a reference signal.
Given a set of loudspeakers producing a loudness $L_i$  at the listening position, the loudness compensation $\Delta L_i$ for each loudspeaker is
\begin{equation}
    \Delta L_i = L_\text{ref} - L_i, 
\end{equation}
where $L_\text{ref}$ is a pre-established reference level. The loudness compensation gains are given by $10^{ \Delta L_i/20}$.

To maintain loudness consistency, the gains $g_i$ produced by a panner are usually normalized so that the loudness of the phantom sound source is equal to the loudness of the corresponding sound source when emanating only from a single loudspeaker. For loudness-matched setups, this requires the following relationship be satisfied:
\begin{equation}\label{eq:gainnorm}
    \left(\sum_j \left|g_j\right|^p \right)^{1/p}=1, 
\end{equation}
with $p$ usually between 1 and 2. The common sine/cosine pairwise panning law is an example which satisfies the above condition for $p=2$. However, any panning law can meet this requirement through normalization. %

Loudspeaker time alignment consists of adding time delays to the closer loudspeakers so that all loudspeaker signals arrive at the listening position at the same time. The delays $\Delta t_i$ applied to each loudspeaker are
\begin{equation}
    \Delta t_i = (d_{\text{ref}} - d_i)/c,
\end{equation}
where $c$ is the speed of sound and $d_\text{ref}$ is a reference distance, usually the distance to the most distant loudspeaker.

\section{Improved panning on non-equidistant loudspeakers}\label{sec:DSC}

As mentioned in the introduction, we observed that the phantom source is systematically pulled towards the closest loudspeakers when using the full response compensation 
approach outlined in Sect.~\ref{sec:SOTACalib}. Here we propose an alternate procedure that restores the phantom source to its intended position by matching the direct sound from each loudspeaker and preserving the correct loudness by matching levels derived from the full response.

\subsection{Improved phantom source location: direct sound compensation (DSC)}

\begin{figure*}[htbp]
\begin{center}
\includegraphics[width=0.95\textwidth]{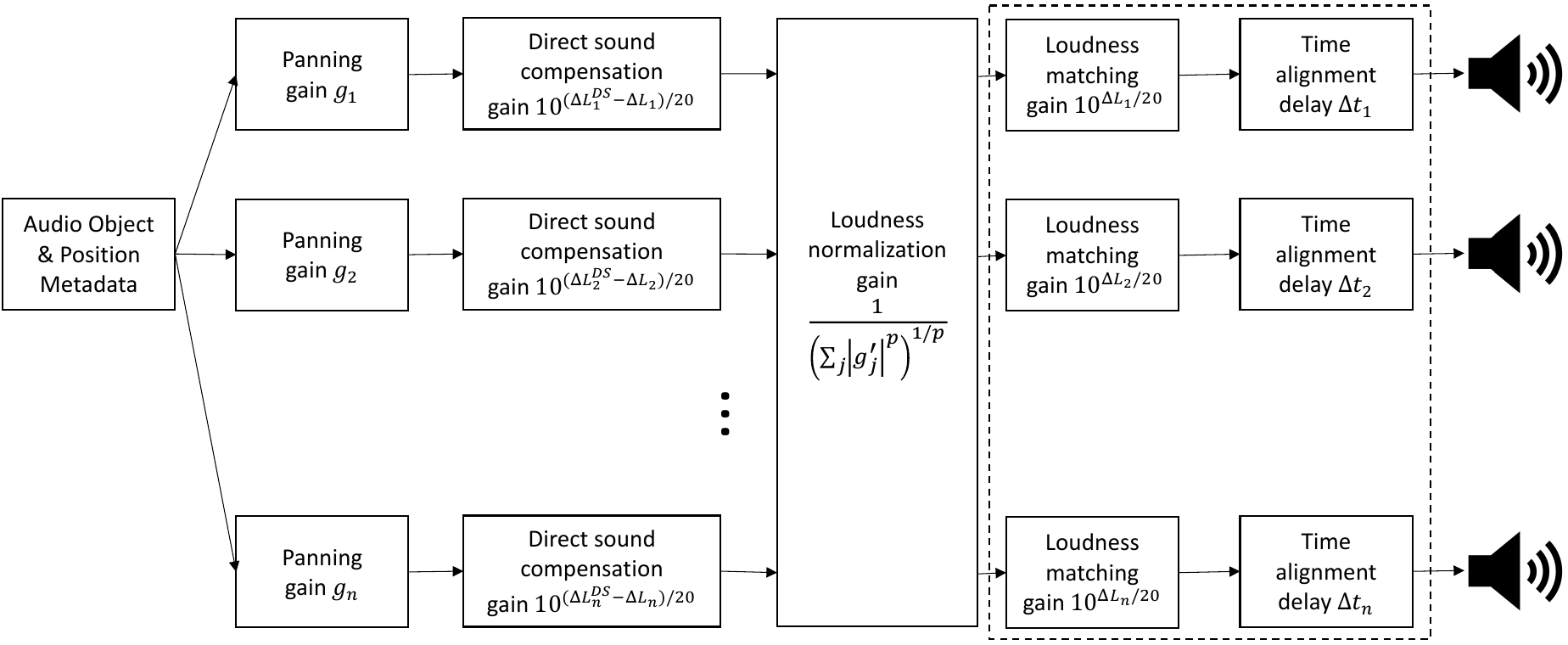}
\caption{System diagram of a panning algorithm enhanced by direct sound compensation, followed by full response loudness compensation and time alignment (dotted box).}%
\label{fg:DSC_System}
\end{center}
\end{figure*}

Given a set of loudspeakers whose direct sound is characterized by a level $L_i^\DS$ as measured in decibels from the listener position, the direct sound compensation for each loudspeaker  $\Delta L_i$ is
\begin{equation}
    \Delta L_i^\DS =L_\text{ref} ^\DS  -L_i^\DS,
\end{equation}
where $L_\text{ref} ^\DS$ is a reference direct sound level. The direct-sound compensation gains are $10^{ \Delta L_i^\DS/20} $.

We assume that the loudspeaker calibration according to the full response compensation procedure outlined in Sec.~\ref{sec:SOTACalib} is already in place.
To preserve the correct phantom source locations, the direct sound compensation needs to be applied to the gains and the effect of loudness compensation needs to be undone. Therefore, the panning gains $g_i$ coming from the amplitude panning algorithm are modified as follows:
\begin{equation}\label{eq:DSCgain}
    g_i \to g_i'=10^{ (\Delta L_i^\DS - \Delta L_i)  / 20} g_i.
\end{equation}

\subsection{Loudness correction}

The application of~\eqref{eq:DSCgain} will lead to phantom source images in their correct location, but the loudness of each one of the phantom sources will generally not be correct as the perception of loudness is governed by the level of the entire room response, and not only by the direct sound. 
To recover the correct loudness of the phantom sound sources, gains $g_i'$  coming from the process of direct sound compensation are normalized to meet the condition in~\eqref{eq:gainnorm}:
\begin{equation}\label{eq:gainrenorm}
    g_i'\to g_i''=\frac{g_i'}{\left(\sum_j \left|g_j'  \right|^p \right)^{1/p} }.
\end{equation}
The complete system, a combination of the full response compensation approach with the additional direct sound compensation gain per source object is depicted in Fig.~\ref{fg:DSC_System}.

Combining the gains stages from~\eqref{eq:DSCgain} and \eqref{eq:gainrenorm} along with the full response loudness compensation gains $10^{ \Delta L_i/20}$, the combined gains $G_i$ for a source fed to each loudspeaker are
\begin{equation}\label{eq:totalgain}
    G_i=\frac{10^{ \Delta L_i^\DS /20} g_i}{\left(\sum_j\left|10^{(\Delta L_j^\DS -\Delta L_j)/20} g_j \right|^p \right)^{1/p} } .
\end{equation}

Should the method outlined here be applied to a loudspeaker setup calibrated in a different way than the state of the art FRC procedure, the specific details in Fig.~\ref{fg:DSC_System}, as well as~\eqref{eq:DSCgain} and \eqref{eq:gainrenorm}, would change, but~\eqref{eq:totalgain} above would still be valid.  

\subsection{Practical implementation}

From~\eqref{eq:totalgain} the final panning gains are clearly dependent on the specifics of the loudspeaker layout, but more critically they are dependent in a manner that varies with phantom source location. This may be appreciated by noting that the denominator of~\eqref{eq:totalgain} is a function of all the unmodified amplitude panning gains $g_i$ across all loudspeakers and will therefore in general be different for different phantom source locations. 
 
As such, a practical implementation requires a render-at-playback-time approach, where the panning gains of each source are applied independently based on the actual loudspeaker layout before mixing together into loudspeaker feeds. This allows for the accounting of direct sound and overall level differences on a per-source basis.  %
This approach works naturally with object-based audio formats but can also be applied to pre-rendered channel-based formats by treating each channel as a "static object" with an assumed canonical playback position. %

This paper presents a broadband analysis and compensation of direct sound and overall loudness. All considerations can be extended to frequency dependent, narrowband calibration based on measurements in the listening room.

\section{Experimental methods}\label{sec:experiment} 

To formally confirm the theoretical and practical findings, a two-part listening test was conducted, isolating the audio attributes of interest respectively: one part focused on the spatial location of phantom sound sources described in Sec.~\ref{sec:LocalizationTest}; the second part targeted the validation of applied loudness correction described in Sec.~\ref{sec:LoudnessTest}. 

The physical audio system was shared between the two experiments and was set up in an acoustically untreated room, matching typical living room conditions. It consisted of two stereo setups each with loudspeakers at $30^\circ$ and $-30^\circ$. One setup had the two loudspeakers placed equidistant at \SI{300}{\cm} with a height of \SI{120}{\cm}. %
The other one had the left loudspeaker at half the distance (\SI{150}{\cm}) of the right one (\SI{300}{\cm}). Both loudspeakers for this non-equidistant setup were at a height of \SI{104}{\cm}. 
A small loudspeaker model (Genelec 8020) was chosen to minimize acoustic impact in the form of occlusion and scattering from the lower, closer loudspeaker on the one behind it. The average ear height of the seated participants was \SI{112}{\cm}, in the middle between the two systems, ensuring an undisturbed acoustic path of both loudspeaker setups to the listener.

\begin{figure}[htbp]
  \centering
  \begin{subfigure}[b]{0.95\columnwidth}
     \begin{overpic}[width=0.95\columnwidth]{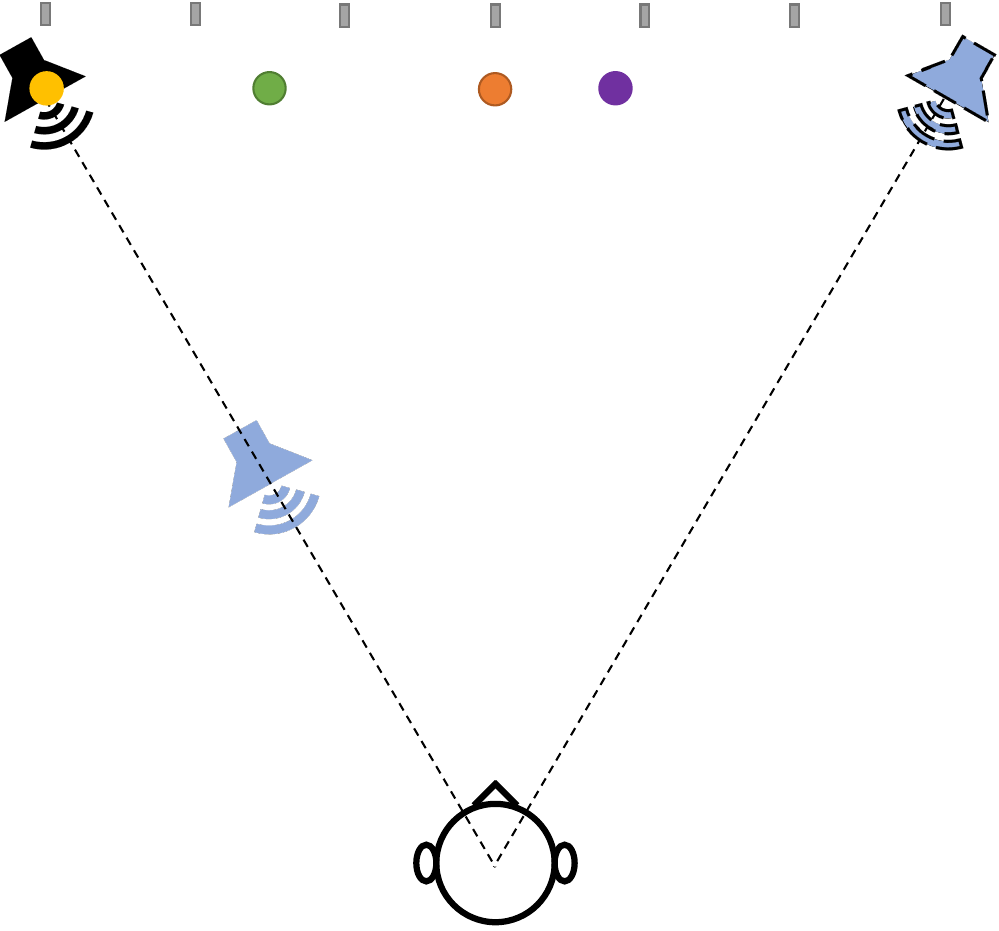} %
        \put (20,25) {\SI{150}{\cm}}
        \put (85,60) {\SI{300}{\cm}}
        \put (0,60) {\SI{300}{\cm}}
        \put (10,82.75) {$30^\circ$}
        \put (29.5,82.75) {$15^\circ$}
        \put (52,82.75) {$0^\circ$}
        \put (63.5,82.75) {$-8^\circ$}
    \end{overpic}
    \hspace{0.5cm}
  \end{subfigure}

  \begin{subfigure}[b]{0.95\columnwidth}
    \begin{overpic}[width=0.95\columnwidth]{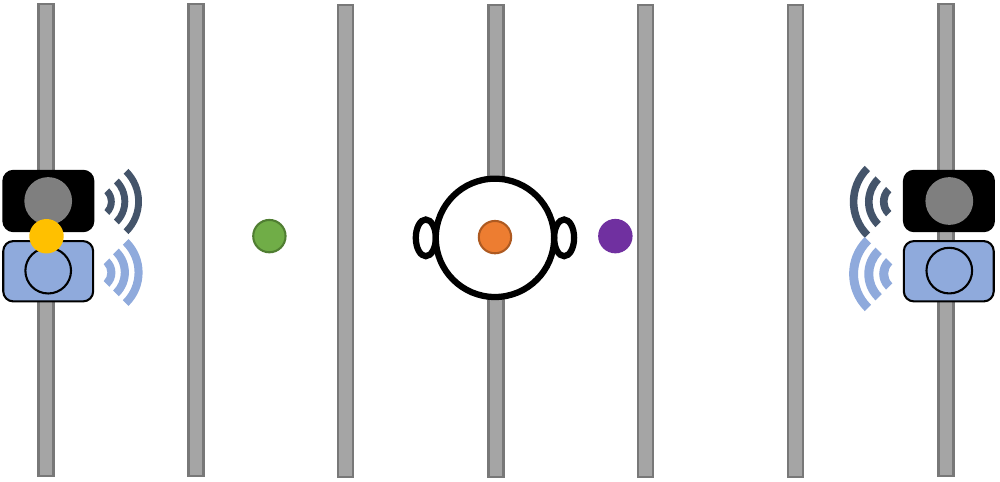}
    \end{overpic}
    \hspace{0.5cm}
  \end{subfigure}
  
  \begin{subfigure}[b]{0.95\columnwidth}
    \includegraphics[width=0.95\columnwidth]{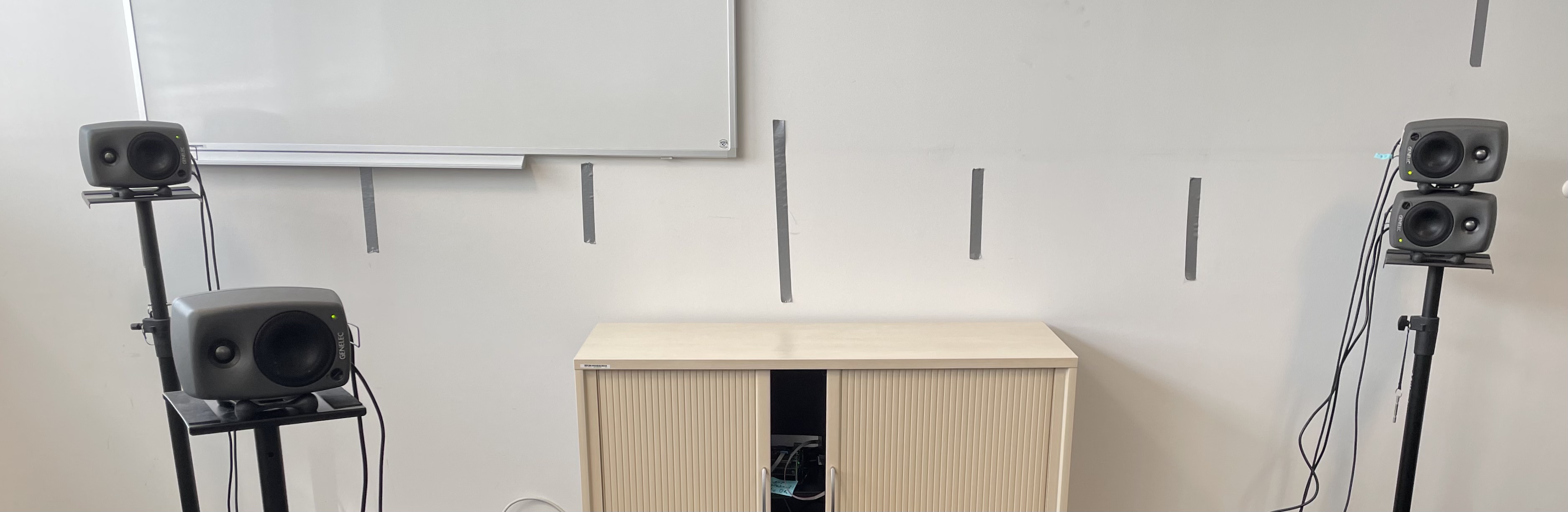}
  \end{subfigure}
  \caption{A schematic top and front view of the listening test setup along with a picture of the actual setup. Four dots indicate the intended phantom sources angle at $30^\circ$, $15^\circ$, $0^\circ$ and $-8^\circ$. 10 degree markers help the participants connect reality to the listening test interface.}
  \label{fig:ListeningTestSetup}
\end{figure}

Fig.~\ref{fig:ListeningTestSetup} shows a schematic view of the listening test setup along with a picture of the actual setup.
Loudspeakers are delay and level aligned according to the FRC calibration procedure based on measured impulse responses. The corresponding IRs, which were also analyzed by FDT to ascertain the direct sound levels, can be seen in Fig.~\ref{fg:IRs}. These direct sound levels matched the inverse square law \eqref{eq:Lds}.
The listening test was realized using the webMUSHRA software \cite{schoeffler2018}.

There were 16 participants (13 male, 3 female) with an average age of 39.4 years. In a questionnaire 56\% stated that they are audio professionals, 43\% had past listening test experience and 19\% claimed to be expert spatial audio listeners. %

\begin{figure}[htbp]
\begin{center}
\includegraphics[width=0.95\columnwidth]{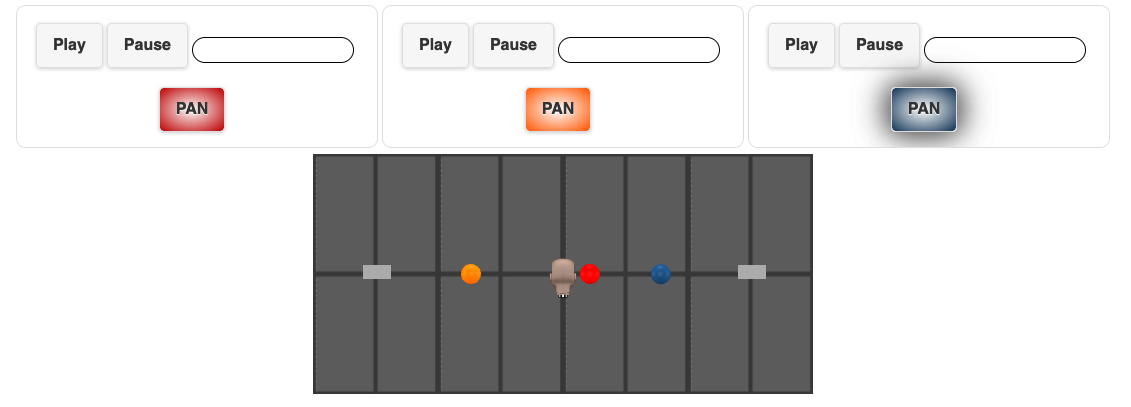}
\caption{Listening test interface used the localization experiment. The intended angle is shared among the three conditions per page. REF, FRC and DSC systems are rated simultaneously.}
\label{fg:WebmushraInterface}
\end{center}
\end{figure}
\subsection{Localization test}\label{sec:LocalizationTest}

In the first listening test, participants were asked to evaluate the perceived angle of phantom sound sources. As shown in Fig.~\ref{fg:WebmushraInterface} three conditions were presented on each page of the listening test software. Each of these conditions used the same mono source content panned to an intended angle using three different panning approaches. For all three the underlying panning law was sin/cos panning. Intended source angles were  30\degree{}, 15\degree{}, 0\degree{} and $-8$\degree{}.  The REF condition utilized the equidistant loudspeakers. The FRC condition refers to the non-equidistant loudspeakers which are delay and level aligned (see Sec.~\ref{sec:SOTACalib}). DSC refers to the panning on the same system according to the methodology described in Sec.~\ref{sec:DSC}.
The mono source content was a selection of a pop song, pink noise bursts, female speech, drums, and harpsichord samples.
The UI position of each stimulus was initialized to a random position; similarly the order of all stimuli was randomized. The participants were instructed to switch between the three conditions on each page and drag and drop little spheres to the desired positions indicating the perceived azimuth location of the phantom sound sources. $10^\circ$ step markers on the wall of the room matched identical indicators in the listening test software user interface and helped the listeners to connect it to reality.

Five participants were excluded from the localization test. Four of them were excluded because in more than 15\% of the cases they reported a hard panning to the left loudspeaker  (30\degree{}) anchor as being located at less than 15\degree{}. Another participant was excluded due to inconsistent reporting.

\subsection{Loudness test}\label{sec:LoudnessTest}
To validate accurate loudness correction for phantom sound sources, listeners were asked to participate in a second part of the listening test.
The utilized methodology was adapted from the loudness validation test proposed in \cite{berendes2022}.
The standardized ITU BS.1534 MUSHRA \cite{MUSHRA} interface was used, where the explicit and hidden reference was a panned source on the symmetric loudspeaker layout (REF).
The participants were asked to evaluate the loudness of the same phantom source panned on the non-equidistant loudspeaker setup with respect to their similarity to the reference purely with respect to loudness. %
Two variants of DSC panned sources were presented, depending on whether direct sound compensation included the loudness correction in~\eqref{eq:gainrenorm}, or not: DSC LC, with loudness correction, and DSC NO LC, without it. Furthermore, an anchor in the form of a scaled reference at \SI{-10}{dB} was added (ANCH). Listeners provided a rating according to the MUSHRA scale with verbal anchors of \textit{bad, poor, fair, good} and \textit{excellent}.
Phantom sound sources were panned to 30\degree{}, 15\degree{}, 0\degree{} with the same mono content selection from the previous part of the test. To shorten the length of the test $-8^\circ$ was left out since the smallest differences were expected for it. 

One participant was excluded from the second test on the basis of evaluating more than 15\% of the hidden reference cases with less than 90 points. 

\section{Experimental results}\label{sec:results}

\begin{figure}[t]
\begin{center}
\includegraphics[width=0.95\columnwidth]{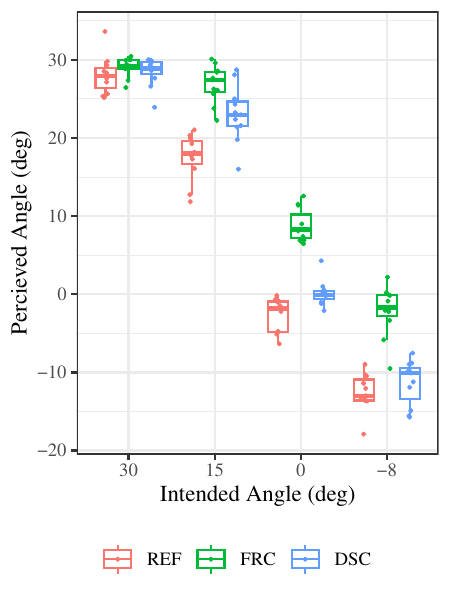}
\caption{Localization test: perceived angular locations, as a function of the four intended angles [30\degree{}\ (i), 15\degree{}\ (ii), 0\degree{}\ (iii), $-8$\degree{}\ (iv)] and the test condition (REF, FRC, DSC).  Dots represent the result of each one of the participants, averaged over all 5 contents items, and the box plots show the corresponding median values and interquartile range.}
\label{fg:3RawRatings}
\end{center}
\end{figure}

\begin{figure}[t]
\begin{center}
\includegraphics[width=0.95\columnwidth]{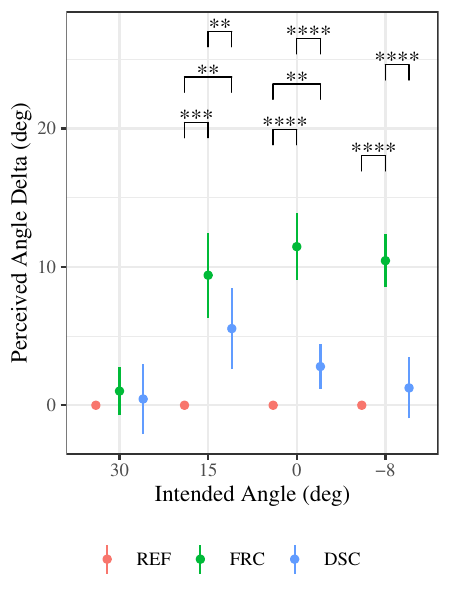}
\caption{Localization test: Mean delta perceived angular positions relative to the reference. Dots represent the mean values and bars the confidence intervals of the mean (95\% CL). The stars in the plot indicate statistically significant $t$-tests (adjusted for multiple comparisons). One star (*) denotes $p<.05$, two stars (**) denote $p < .01$, three stars (***) denote $p < .001$, and four stars (****) denote $p <  10^{-4}$.}
\label{fg:7DeltaAzimuthResults}
\end{center}
\end{figure}

\begin{figure}[t]
\begin{center}
\includegraphics[width=0.9\columnwidth]{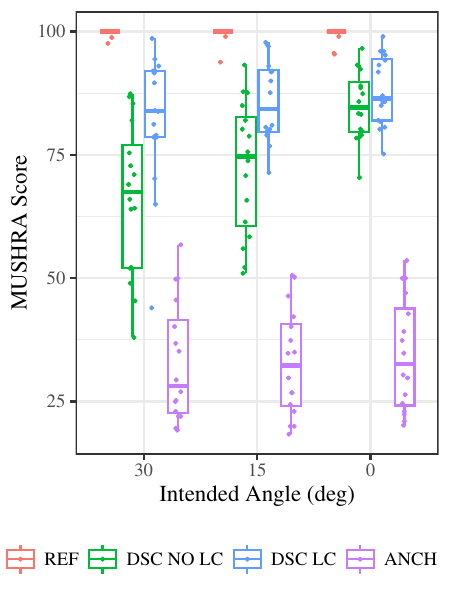}
\caption{Loudness test: MUSHRA score as a function of the four intended angles [30\degree{}\ (i), 15\degree{}\ (ii), 0\degree{}\ (iii)] and the test condition (REF, DSC NO LC, DSC LC, ANCH).  Dots represent the result of each one of the participants, averaged over all 5 content items, and box plots show the corresponding median values and interquartile range.}
\label{fg:3LoudnessRawRatings}
\end{center}
\end{figure}

\begin{figure}[t]
\begin{center}
\includegraphics[width=0.9\columnwidth]{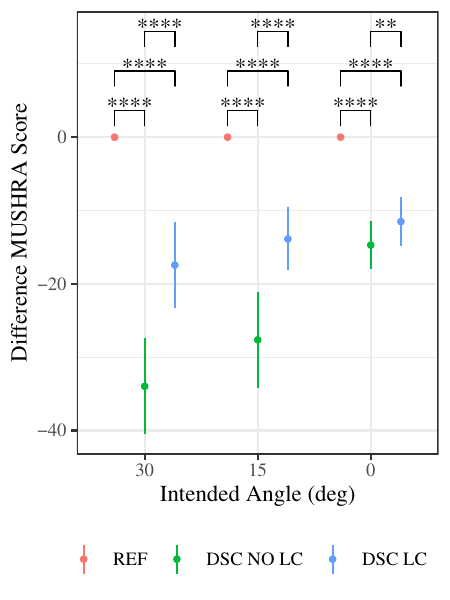}
\caption{Loudness test: Differential mean MUSHRA scores relative to the reference as a function of the three intended angles and the test condition.  Dots represent the mean values and bars the confidence intervals of the mean (95\% CL). Results shown have undergone a standarization process betweeen the different participants (see main text). See caption of Fig.~\ref{fg:7DeltaAzimuthResults} for the meaning of the significance stars.}
\label{fg:7LoudnessDeltarating_scoreResults}
\end{center}
\end{figure}

The statistical analysis follows the general guidelines in ITU-R BS.1534 \cite{MUSHRA}, and was done using the rstatix package in R \cite{rstatix, r}.

\subsection{Localization test}

Initially, the normality of the data was examined by means of a QQ plot, which revealed no apparent deviations from normality.
A 3-way repeated measures ANOVA was conducted to examine whether the perceived angular positions were dependent on the test content. No significant interaction was revealed [$F(8, 80) = 0.8$, $p=.6$].

Subsequently, results were averaged over the  different source content items. The resulting data distribution is shown in Fig.~\ref{fg:3RawRatings}, as a function of the three test conditions (REF, FRC, DSC) and the four panning angles [30\degree{}\ (i), 15\degree{}\ (ii), 0\degree{}\ (iii), $-8\degree{}$\ (iv)]. The median perceived positions for the symmetric reference system were 28\degree{} (i), 18\degree{} (ii), $-2\degree{}$ (iii), and $-13\degree{}$ (iv), showing a slight displacement  from their nominal positions.

A 2-way repeated measures ANOVA was performed to examine the effects of the test condition and intended angle on the results. The ANOVA confirmed significant main effects for the test condition [$F(2,20)=138.7$, $p=2\times 10^{-12}$], as well as a significant interaction between the test condition and intended angle [$F(3.0,29.5)=17.0$, $p=1\times 10^{-6}$].

To further investigate the differences between angles and the three test conditions, multiple paired $t$-tests were conducted. We utilized the Benjamini–Hochberg method to account for multiple comparisons \cite{MUSHRA}; all stated $p$-values are already adjusted for this correction. Refer to Fig.~\ref{fg:7DeltaAzimuthResults} for a depiction of the perceived angle deltas with respect to the reference and the results of the paired $t$-tests. At 30\degree{} (hard panning to the left loudspeaker), all panning methods were statistically indistinguishable from one another ($p \geq .4$). For the remaining phantom source positions, the average FRC results exhibited a consistent displacement of 9 to 11 degrees towards the closest loudspeaker, with these differences being  significant in all cases ($p \leq 1 \times 10^{-4}$). The average DSC results were much closer to the reference, but still displayed a slight displacement towards the closest loudspeaker: 6\degree{} (ii), 3\degree{} (iii), and 1\degree{} (iv). The differences were  significant in cases (ii) and (iii) ($p\leq .002$), but not in case (iv) ($p = .2$).

\subsection{Loudness test}

Initially we examined whether the test results of the loudness validation test were dependent on the test content. A 3-way repeated measures ANOVA revealed no significant interaction between the test condition and the content item [$F(4.4, 38) = 2.3$, $p=.07$]. 

Subsequently, results were averaged over the  different source content items. The resulting data distribution is shown in Fig.~\ref{fg:3LoudnessRawRatings}, as a function of the four test conditions (REF, DSC NO LC, DSC LC, ANCH) and the three panning angles [30\degree{}\ (i), 15\degree{}\ (ii), 0\degree{}\ (iii)].
The DSC LC condition always scores in the excellent range (above 80 MUSHRA points). The DSC NO LC condition scores systematically below DSC LC, the difference being greater for panning angles closer to the left loudspeaker.

The QQ plot initially indicated moderate deviations from normality, which were determined to be a result of participants rating content on differing scales. To address this, a data normalization procedure was implemented. Specifically, each participant's result was standardized to have zero mean and unit variance. The MUSHRA scale was then restored by multiplying the standarized results by the global variance and adding the global mean. Following this procedure, the QQ plot no longer indicated evident deviations from normality of the data. The anchor was discarded from the subsequent analysis.

We conducted a 2-way repeated measures ANOVA to examine the effects of test condition and the intended angle on the results. The ANOVA analysis confirmed a significant main effect for the test condition [$F(1.4,  20.8) = 90.3$, $p = 6\times 10^{-10}$] and significant interaction between test condition and intended angle [$F(2.7, 40.0) = 28.1$,  $p = 2 \times 10^{-9}$].
 
 A subsequent post-hoc analysis was conducted, in the form of multiple paired $t$-tests between the different test conditions (see Fig.~\ref{fg:7LoudnessDeltarating_scoreResults}). Again, Benjamini–Hochberg correction for multiple comparisons \cite{MUSHRA} was applied. Analysis showed that without loudness correction, scores are  on average 33 (i), 27 (ii), and 14 (iii) MUSHRA points lower than the reference on average. With loudness correction, this difference is reduced to 16 (i), 13 (ii), and 12 (iii) MUSHRA points. All mutual comparisons are significant ($p \leq .005$).

\section{Discussion}\label{sec:summary}
The results of the experimental tests show that the common practice to time and level align loudspeakers is insufficient when dealing with non-equidistant loudspeakers, as the phantom source for the FRC system is consistently skewed towards a closer loudspeaker. The average perceived angle delta of about 10\degree{} across all angles under test is high and would result in a significantly impaired playback performance. 
In all tested cases, using the proposed DSC approach significantly improves the delta angle towards the intended panning position. 
It is noteworthy that, according to the experiment, DSC performs particularly well in the area in front of the listener, where the human hearing is most sensitive to angular changes.

It is worth mentioning that at the largest examined panning angle (15\degree) the experiment still showed a relatively high bias towards the closer loudspeaker (6\degree{}). While it is possible that the calculated compensation gain was not totally accurate, it could be conceivable that visual cues of the close loudspeaker pull the rating towards it as the intended panning position comes close to it. After all, phantom source localization is a complicated task affected by multi-sensory factors.

The results of the loudness validation test are in agreement with our hypothesis that the loudness of phantom sources is not defined by the direct sound, but by the full loudspeaker and room response. Across all tested angles, the DSC panning that was loudness compensated according to the full response, got a mean rating in the \textit{excellent} range of the MUSHRA scale, in all cases significantly better than the non-loudness compensated version of DSC. The dependence on angle of the ratings for the non-loudness compensated condition nicely match the calculated diminishing dB value as the phantom source is panned further and further away from the closer loudspeaker.
It is unsurprising that listeners rated differences between the loudness compensated DSC and the reference system. Those can mainly be attributed to other differences in the systems' characteristics such as direct-to-reverberant ratio or spatial characteristics of close loudspeakers which might not have been  fully ignored by the listeners, though instructed to do so.

While the formal listening test considered a single phantom source on a stereo layout, typical multimedia content contains many panned objects and the restoring effect to the intended positions using DSC accumulates. Listeners participating in informal listening using a DSC enabled object renderer and multiple loudspeakers reported not only on the restoration of the overall balance, which is otherwise heavily skewed towards close loudspeakers, but also commented on the vastly improved clarity of the mix. These effects were reported to positively affect the rendering, even when the difference in loudspeaker distance was not as substantial as in the presented listening test. Especially in the context of object-based audio and flexible rendering engines at playback time, this approach is a notable step forward towards a faithful representation of the artistic intent in the consumer environment. As object based content makes its way into more and more playback systems like living rooms or cars, the typical loudspeaker setup will be increasingly in-homogeneous and non-equidistant. Instead of forcing consumers to place loudspeakers in canonical positions, the system should be able to adapt. In this paper we have layed out that, as a consequence of the precedence effect, any panning algorithm and renderer will benefit from taking into account the importance of the relative direct sound.

\bibliographystyle{jaes}

\bibliography{refs}

\end{document}